\documentclass[final,5p,times,twocolumn]{elsarticle}

\usepackage[T1]{fontenc}
\usepackage[utf8]{inputenc}

\usepackage{amsmath}
\usepackage{amsfonts}
\usepackage{amssymb}
\usepackage{amsxtra}
\usepackage{array}
\usepackage{color}
\usepackage{dcolumn}
\usepackage{graphicx}
\usepackage{hepunits}
\usepackage{units}
\usepackage{xspace}
\usepackage{CJKutf8}

\definecolor{purple}{rgb}{0.5,0,0.5}
\definecolor{blue}{rgb}{0.0,0,0.9}
\definecolor{prdblue}{rgb}{0.133,0.118,0.498}
\usepackage[colorlinks=true, pdfstartview=FitV, linkcolor=prdblue, citecolor= prdblue, urlcolor=prdblue]{hyperref}

\usepackage[mathscr,scaled=1.15]{urwchancal}
\DeclareFontFamily{OT1}{pzc}{}
\DeclareFontShape{OT1}{pzc}{m}{it}%
{<-> s * [1.15] pzcmi7t}{}
\DeclareMathAlphabet{\mathpzc}{OT1}{pzc}{m}{it}

\biboptions{sort&compress}

\journal{Physics Letters B}

\begin{document}
\begin{CJK}{UTF8}{song}

\begin{frontmatter}

\title{$\,$\\[-7ex]\hspace*{\fill}{\normalsize{\sf\emph{Preprint no}. NJU-INP 056/22}}\\[1ex]
Proton and pion distribution functions in counterpoint}

\author[NJU,INP,NJTU]{Ya Lu}

\author[NKU]{Lei Chang}

\author[UGranada,UNAM]{Kh\'epani Raya}

\author[NJU,INP]{Craig D. Roberts}

\author[UHe]{Jos\'e Rodr\'{\i}guez-Quintero}

%
\address[NJU]{
School of Physics, Nanjing University, Nanjing, Jiangsu 210093, China}
\address[INP]{
Institute for Nonperturbative Physics, Nanjing University, Nanjing, Jiangsu 210093, China}

\address[NJTU]{
Department of Physics, Nanjing Tech University, Nanjing 211816, China}
\address[NKU]{
School of Physics, Nankai University, Tianjin 300071, China}

\address[UGranada]{Departamento de F\'isica Te\'orica y del Cosmos, Universidad de Granada, E-18071, Granada, Spain}

\address[UNAM]{Instituto de Ciencias Nucleares, Universidad Nacional Aut\'onoma de M\'exico, Apartado Postal 70-543, CDMX 04510, M\'exico}

\address[UHe]{
Department of Integrated Sciences and Center for Advanced Studies in Physics, Mathematics and Computation, 
University of Huelva, E-21071 Huelva, Spain\\[1ex]
%
\href{mailto:luya@nju.edu.cn}{luya@nju.edu.cn} (Y. Lu);
\href{mailto:leichang@nankai.edu.cn}{leichang@nankai.edu.cn} (L. Chang);
\href{mailto:khepani@ugr.es}{khepani@ugr.es} (K. Raya);\\
\href{mailto:cdroberts@nju.edu.cn}{cdroberts@nju.edu.cn} (C. D. Roberts);
\href{mailto:jose.rodriguez@dfaie.uhu.es}{jose.rodriguez@dfaie.uhu.es} (J. Rodr{\'{\i}}guez-Quintero)
\\[1ex]
Date: 2022 03 01\\[-6ex]
}

\begin{abstract}
Working with proton and pion valence distribution functions (DFs) determined consistently at the same, unique hadron scale and exploiting the possibility that there is an effective charge which defines an evolution scheme for DFs that is all-orders exact, we obtain a unified body of predictions for all proton and pion DFs -- valence, glue, and four-flavour-separated sea.  Whilst the hadron light-front momentum fractions carried by identifiable parton classes are the same for the proton and pion at any scale, the pointwise behaviour of the DFs is strongly hadron-dependent.  All calculated distributions comply with quantum chromodynamics constraints on low- and high-$x$ scaling behaviour and, owing to emergent hadron mass, pion DFs are the most dilated.  These results aid in elucidating the sources of similarities and differences between proton and pion structure.
\end{abstract}

\begin{keyword}
continuum Schwinger function methods \sep
emergence of mass \sep
pion structure \sep
proton structure \sep
parton distributions \sep
strong interactions in the standard model of particle physics
\end{keyword}

\end{frontmatter}
\end{CJK}

\begin{figure}[t]
\begin{minipage}{0.5\textwidth}
\begin{tabular}{cc}
{\sf A} \hspace*{0.4\textwidth} & {\sf B} \hspace*{0.4\textwidth}\\[-0.5ex]
\includegraphics[width=0.45\textwidth]{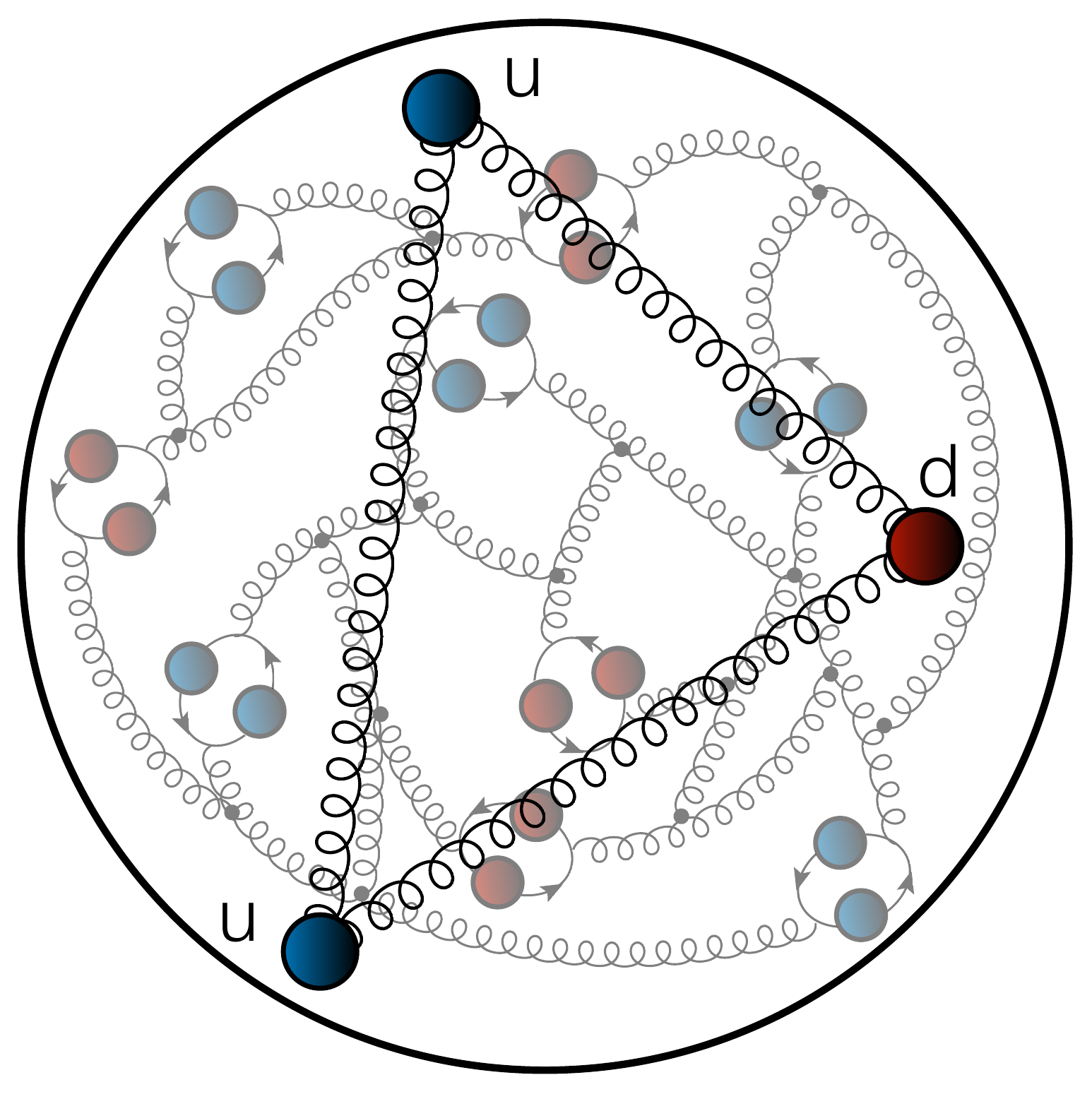}&
\includegraphics[width=0.45\textwidth]{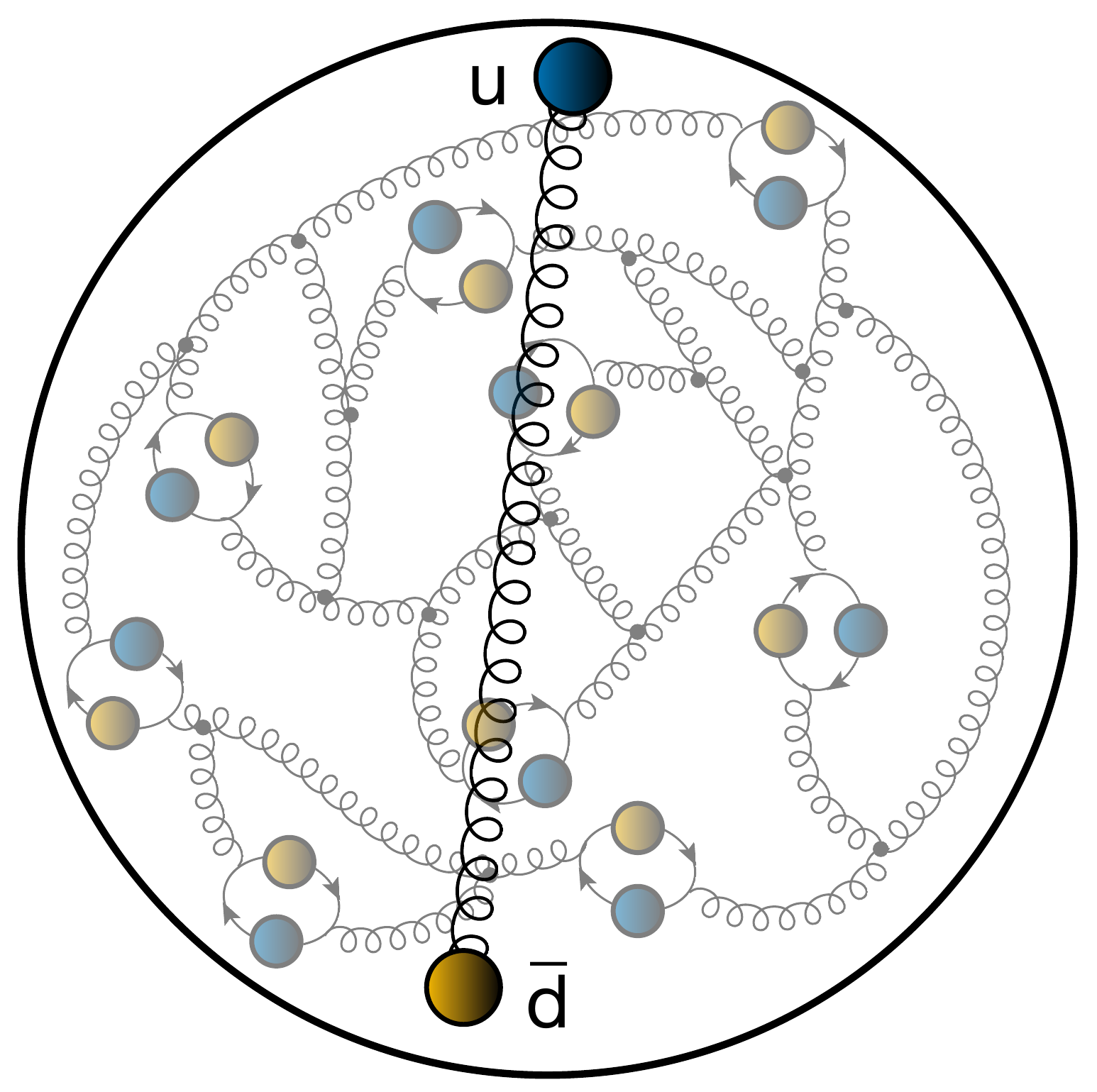}
\end{tabular}
\end{minipage}
\caption{\label{ImageProtonPion}
\emph{Left panel}\,--\,{\sf A}.
In terms of QCD's Lagrangian quanta, the proton, $p$, contains two valence up ($u$) quarks and one valence down ($d$) quark; and also infinitely many gluons and sea quarks, drawn here as ``springs'' and closed loops, respectively.  The neutron, as the proton's isospin partner, is defined by one $u$ and two $d$ valence quarks.
\emph{Right panel}\,--\,{\sf B}. The pion, $\pi^+$, contains one valence $u$-quark, one valence $\bar d$-quark, and, akin to the proton, infinitely many gluons and sea quarks.  (In terms of valence quarks, $\pi^- \sim d\bar u $ and $\pi^0 \sim u\bar u - d\bar d$.)
}
\end{figure}

\noindent\textbf{1.$\;$Introduction}.
Protons, neutrons, and pions are amongst the most fundamental entities in Nature.  From many perspectives, these hadrons are the primary components of atomic nuclei; yet, within the standard model of particle physics, they are bound-states, built from the gluon and quark parton fields used to express the Lagrangian of quantum chromodynamics (QCD) \cite{Marciano:1979wa}.  The light up ($u$) and down ($d$) quarks are key here.  They were the first quarks discovered \cite{Riordan:1992hr} and provide the seeds for the proton, $p$, which is comprised of one valence $d$ and two valence $u$ quarks, hence a definitive baryon, and the pions, which, considering the positive charge state, $\pi^+$, is constituted from one valence $u$ quark and one valence $\bar d$ quark ($d$ antiquark) -- definitively, a meson.
However, as highlighted by Fig.\,\ref{ImageProtonPion}, valence quark partons are only part of the explanation for proton and pion structure.   Owing to the character of strong interactions in the standard model, the valence parton quanta are embedded in a dense medium of gluons and sea quarks of their own making \cite{Brodsky:2012ku}.
Viewed from this position, the proton and pion each contain an enumerably infinite number of QCD's Lagrangian quanta; and ever since the formulation of QCD, physics has sought to measure and understand the distributions of these quanta throughout bound-state volumes \cite{Holt:2010vj, Rojo:2015acz, Hen:2016kwk, Hadjidakis:2018ifr}.

Using any framework that enables the reliable calculation of Poincar\'e-covariant wave functions for the proton and pion, one can obtain the light-front wave functions in terms of which the gluon and quark parton distribution functions (DFs) can rigorously be defined \cite{Brodsky:1979gy, Brodsky:1989pv}.  One-dimensional DFs have been the focus of experiment and theory for more than fifty years \cite{Ellis:1991qj}.  These quantities are probability densities, each of which describes the light-front fraction, $x$, of the bound-state's total momentum carried by the given parton species within the bound-state \cite{Holt:2010vj}.  Today, notwithstanding the enormous expense of time and effort, much must still be learnt before proton and pion structure may be considered understood in terms of DFs; even, most simply, what are the differences, if any, between the distributions of partons within the proton and the pion.

The question of similarity/difference between proton and pion DFs has particular resonance today as science seeks to explain emergent hadron mass (EHM) \cite{Papavassiliou:2015aga, Roberts:2016vyn, horn_tanja_2020_4304547, binosi:trento, Roberts:2020hiw, Krein:2020yor, Roberts:2021nhw, denisov:cern2021}.
Namely, how can a Lagrangian that possesses no mass-scale in the absence of Higgs boson couplings produce both an absolutely stable proton with mass $m_p \approx 1\,$GeV and electric charge radius $r_E \approx 0.84\,$fm, and, simultaneously, a massless pion, the Nambu-Goldstone boson associated with dynamical chiral symmetry breaking (DCSB), of similar size?  And supposing it does, how are these macroscopic differences expressed in the structure of these two bound-states?  Modern experiments \cite{JlabTDIS1, JlabTDIS2, Adams:2018pwt, Aguilar:2019teb, Brodsky:2020vco, Chen:2020ijn, Anderle:2021wcy, Arrington:2021biu, AbdulKhalek:2021gbh, Mokeev:2022xfo}, at upgraded and anticipated facilities, aim to provide data that can be used to answer these and related questions.

Working with QCD, some predictions are available.  For instance, considering DFs measured in processes that do not involve beam or target polarisation, then at some hadron scale, $\zeta_{\cal H}<m_p$, the valence-quark DFs in the proton and pion behave as follows \cite{Brodsky:1994kg, Yuan:2003fs, Cui:2021mom, Cui:2022bxn}:
\begin{subequations}
\label{LargeX}
\begin{align}
{\mathpzc d}^p(x;\zeta_{\cal H}), {\mathpzc u}^p(x;\zeta_{\cal H}) & \stackrel{x\simeq 1}{\propto} (1-x)^3\,,\\
\bar {\mathpzc d}^\pi(x;\zeta_{\cal H}), {\mathpzc u}^\pi(x;\zeta_{\cal H}) & \stackrel{x\simeq 1}{\propto} (1-x)^2\,;\;
\end{align}
\end{subequations}
the exponent on the associated gluon DFs is approximately one unit larger; and that for the sea quark DFs is roughly two units larger.  With increasing scale, $\zeta > \zeta_{\cal H}$, all these exponents increase logarithmically in a manner prescribed by the DGLAP equations \cite{Dokshitzer:1977sg, Gribov:1971zn, Lipatov:1974qm, Altarelli:1977zs}.
However, feeding controversy and leading some to challenge the veracity of QCD \cite{Aicher:2010cb, Cui:2021mom, Cui:2022bxn}, these constraints are typically ignored in fits to the world's deep inelastic scattering data \cite{Ball:2016spl, Hou:2019efy, Bailey:2020ooq, Novikov:2020snp, Barry:2021osv}.  Furthermore, proton and pion data have not been considered simultaneously, largely because pion data are scarce \cite[Table~9.5]{Roberts:2021nhw}.

Against this backdrop, we exploit recent progress made using continuum Schwinger function methods (CSMs) \cite{Cui:2021sfu, Cui:2020tdf, Chang:2021utv, Cui:2021mom, Cui:2022bxn, Chang:2022jri} in developing a unified set of predictions for all proton and pion DFs.  Crucially, their common origin enables meaningful comparisons to be made between them.

\medskip

\noindent\textbf{2.$\;$Hadron scale and DF evolution}.
Using CSMs, the hadron scale, $\zeta_{\cal H}$, is naturally identified with the resolving scale at which dressed valence degrees-of-freedom carry all measurable properties of the hadron, including its light-front momentum \cite{Cui:2021sfu, Cui:2020tdf, Chang:2021utv, Cui:2021mom, Cui:2022bxn, Chang:2022jri}, and $\zeta_{\cal H}$ is the same for both the proton and the pion.
Defining the $n^{\rm th}$ moment of a given DF as ($H=p,\pi$)
\begin{equation}
\langle x^n \rangle_{{\mathpzc p}_H}^\zeta = \int_0^1\,dx\,x^n\,{\mathpzc p}(x;\zeta)\,,
\end{equation}
then this identification of $\zeta_{\cal H}$ entails
\begin{equation}
\label{MomSum}
\langle x \rangle_{{\mathpzc u}_p}^{\zeta_{\cal H}} +
\langle x \rangle_{{\mathpzc d}_p}^{\zeta_{\cal H}} = 1\,,\quad
\langle x \rangle_{{\mathpzc u}_\pi}^{\zeta_{\cal H}} +
\langle x \rangle_{\bar {\mathpzc d}_\pi}^{\zeta_{\cal H}} = 1\,;
\end{equation}
further, that all glue and sea DFs vanish identically at $\zeta_{\cal H}$.

At this point, given hadron scale valence DFs for the proton and pion, then predictions for all DFs at any scale $\zeta > \zeta_{\cal H}$ follow immediately from the following proposition \cite{Cui:2021mom, Cui:2022bxn}:\\[0.4ex]
\hspace*{0.8\parindent}\parbox[t]{0.9\linewidth}{
{\sf P1}: There exists an effective charge, $\alpha_{1\ell}(k^2)$, that, when used to integrate the one-loop perturbative-QCD DGLAP equations, defines an evolution scheme for parton DFs that is all-orders exact.
}

\medskip

\noindent Charges of this type are discussed elsewhere \cite{Grunberg:1982fw, Grunberg:1989xf, Dokshitzer:1998nz}.  They need not be process-independent (PI); hence, not unique.
Nevertheless, a suitable PI charge is available: the coupling discussed in Refs.\,\cite{Cui:2019dwv, Cui:2020tdf, Chang:2021utv} has proved efficacious.  On the other hand, as highlighted elsewhere \cite{Cui:2021mom, Cui:2022bxn}, the pointwise form is largely immaterial.
In being defined by an observable -- in this case, structure functions, each such $\alpha_{1\ell}(k^2)$ is \cite{Deur:2016tte}: consistent with the renormalisation group, independent of renormalisation scheme, everywhere analytic and finite; and, further, provides an infrared completion of any standard perturbative running coupling.
%


Employing this approach and supposing that the evolution kernels are independent of quark mass, explicit solutions of the evolution equations are presented elsewhere \cite[Sec.\,VII]{Raya:2021zrz}.  We now introduce a simple generalisation that expresses salient effects of quark mass dependence in the evolution kernels.  For simplicity of presentation here, we focus on evolution equations for DF moments.

Consider four quark flavours: the two light quarks, ${\mathpzc l}=u,d$, treated as degenerate; strange, $s$; and charm, $c$.  Regarding their dynamically determined mass functions, one may define the following quark infrared masses \cite[Fig.\,2.5]{Roberts:2021nhw} $M_q = \zeta_{\cal H}+\delta_q$, $\delta_l\approx 0$, $\delta_s \approx 0.1\,$GeV, $\delta_c \approx 0.9\,$GeV.
The $\zeta>\zeta_{\cal H}$ scale dependence of the moments of all the hadron's DFs (valence, glue, and singlet \emph{i.e}., $\Sigma_H^q={\mathpzc q}+\bar{\mathpzc q}$, ${\mathpzc q}=u, d, s, c$) are obtained by solving the following set of coupled differential equations, using the nonzero valence DFs as initial values at $\zeta=\zeta_{\cal H}$:
\begin{subequations}
\label{EqEvolution}
\begin{align}
\zeta^2 \frac{d}{d\zeta^2} \langle x^n \rangle_{{\mathpzc q}_H}^\zeta
& = - \frac{\alpha_{1\ell}(\zeta^2)}{4\pi}
\gamma_{qq}^n \langle x^n \rangle_{{\mathpzc q}_H}^\zeta \,,\\
\zeta^2 \frac{d}{d\zeta^2} \langle x^n \rangle_{\Sigma_H^q}^\zeta
& = - \frac{\alpha_{1\ell}(\zeta^2)}{4\pi}
\left[
\gamma_{qq}^n \langle x^n \rangle_{\Sigma_H^q}^\zeta
+ 2 {\cal P}_{qg}^\zeta \gamma_{qg}^n \langle x^n \rangle_{{\mathpzc g}_H}^\zeta
\right]\,, \label{EqCalP}\\
\zeta^2 \frac{d}{d\zeta^2} \langle x^n \rangle_{{\mathpzc g}_H}^\zeta
& = - \frac{\alpha_{1\ell}(\zeta^2)}{4\pi}
\left[\sum_{\mathpzc q}
\gamma_{gq}^n \langle x^n \rangle_{\Sigma_H^q}^\zeta
+ \gamma_{gg}^n \langle x^n \rangle_{{\mathpzc g}_H}^\zeta
\right]\,,  \label{EqCalPc}
\end{align}
\end{subequations}
where $\gamma_{qq}^n$, $\gamma_{qg}^n$, $\gamma_{gq}^n$, $\gamma_{gg}^n$  are anomalous dimensions \cite[Sec.\,VII]{Raya:2021zrz}.
Moments of the sea quark DFs are readily obtained:
\begin{equation}
\langle x^n \rangle_{{\mathpzc S}_H^{q}}^\zeta =
 \langle x^n \rangle_{\Sigma_H^q}^\zeta - \langle x^n \rangle_{{\mathpzc q}_H}^\zeta\,.
\end{equation}
Notably, so long as $\zeta_{\cal H}$ and the evolution equations are the same for a given family of hadrons -- herein, nucleons and pions, then the light-front momentum fractions stored in each parton class are also the same for these kindred hadrons at any scale, \emph{e.g}.:
\begin{equation}
\begin{array}{ll}
\langle x \rangle_{{\mathpzc u}_p+{\mathpzc d}_p}^\zeta = \langle x \rangle_{{\mathpzc u}_\pi+\bar{\mathpzc d}_\pi}^\zeta\,,
&
\langle x \rangle_{{\mathpzc g}_p}^\zeta = \langle x \rangle_{{\mathpzc g}_\pi}^\zeta\,,\\
\langle x \rangle_{\Sigma_p^{u+d}}^\zeta = \langle x \rangle_{\Sigma_\pi^{u+\bar d}}^\zeta\,,\;
&
\langle x \rangle_{\Sigma_p^{s,c}}^\zeta = \langle x \rangle_{\Sigma_\pi^{s,c}}^\zeta\,.
\end{array}
\end{equation}
These equations highlight the relevant four distinct parton classes: collected valence degrees-of-freedom; associated sea quarks; flavour-distinct sea quarks; and glue.

Eq.\,\eqref{EqCalP} features a threshold function ${\cal P}_{qg}^\zeta \sim \theta(\zeta - \delta_q)$.  This factor ensures that a given quark flavour only participates in DF evolution when the resolving energy scale exceeds a value determined by the quark's mass.
%
%
Its effect can be anticipated.
If each quark flavour were light, then all would be emitted with equal probability at any $\zeta>\zeta_{\cal H}$ and evolution would lead to a certain hadron gluon momentum fraction plus a sea-quark fraction shared equally amongst all quark species.
Accounting for mass differences between the quarks, with some heavier than the light-quark threshold, then Eqs.\,\eqref{EqEvolution} entail that at any $\zeta>\zeta_{\cal H}$, the gluon fraction is approximately as it was in the all-light quark case, but the sea-quark momentum fraction is shared amongst the quarks in roughly inverse proportion to their mass.
Reviewing Ref.\,\cite[Sec.\,7.3]{Cui:2020tdf}, one could also introduce a factor multiplying $\gamma_{gq}^n$ in Eq.\,\eqref{EqCalPc} that serves to suppress the emission of gluons by heavier quarks, with a linked momentum-balance correction to $\gamma_{qq}^n$.  However, in the present context, our calculations show this subleading effect to be negligible.

\begin{figure}[!t]
\vspace*{0.5ex}

\leftline{\hspace*{0.5em}{\large{\textsf{A}}}}
\vspace*{-3ex}
\includegraphics[width=0.41\textwidth]{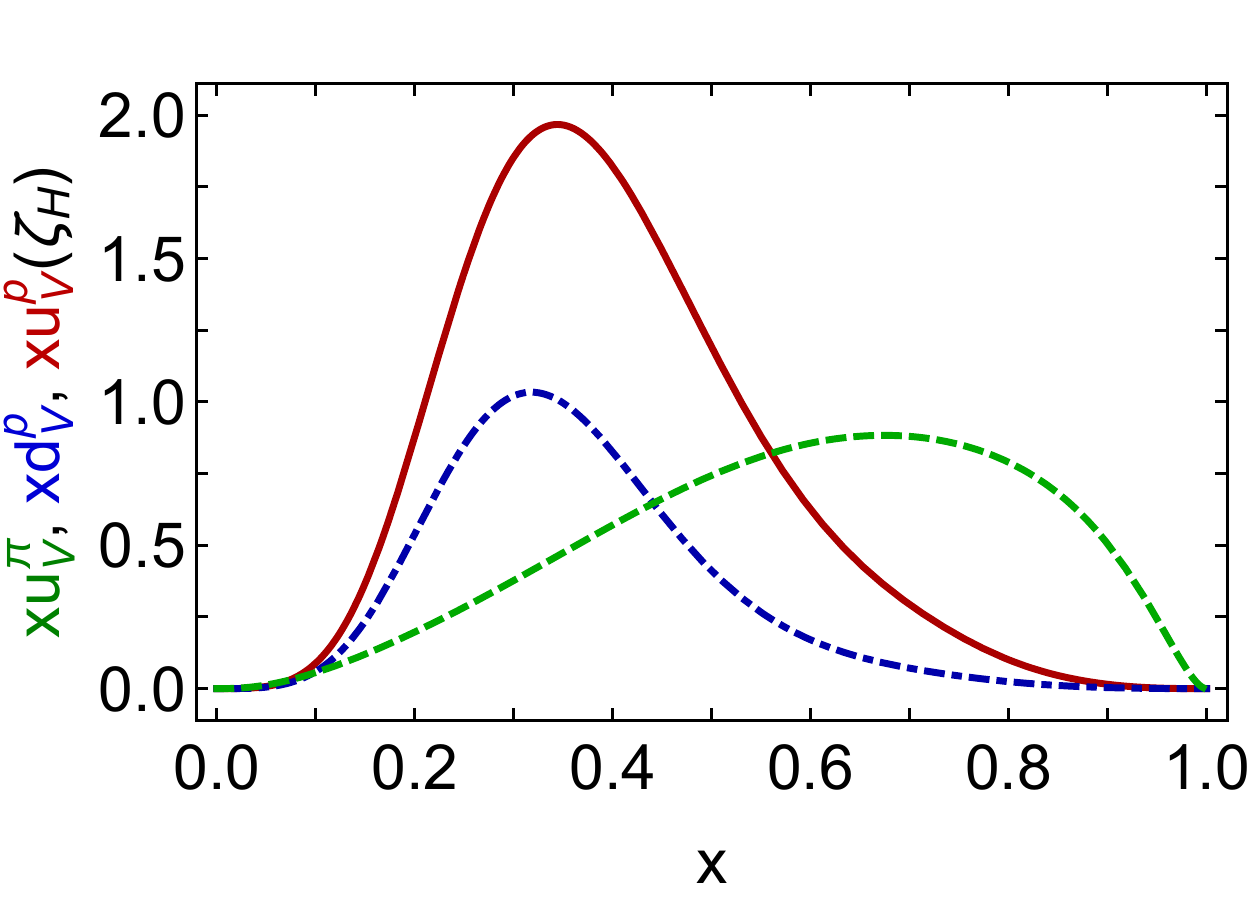}
\vspace*{-1ex}

\leftline{\hspace*{0.5em}{\large{\textsf{B}}}}
\vspace*{-3ex}
\includegraphics[width=0.41\textwidth]{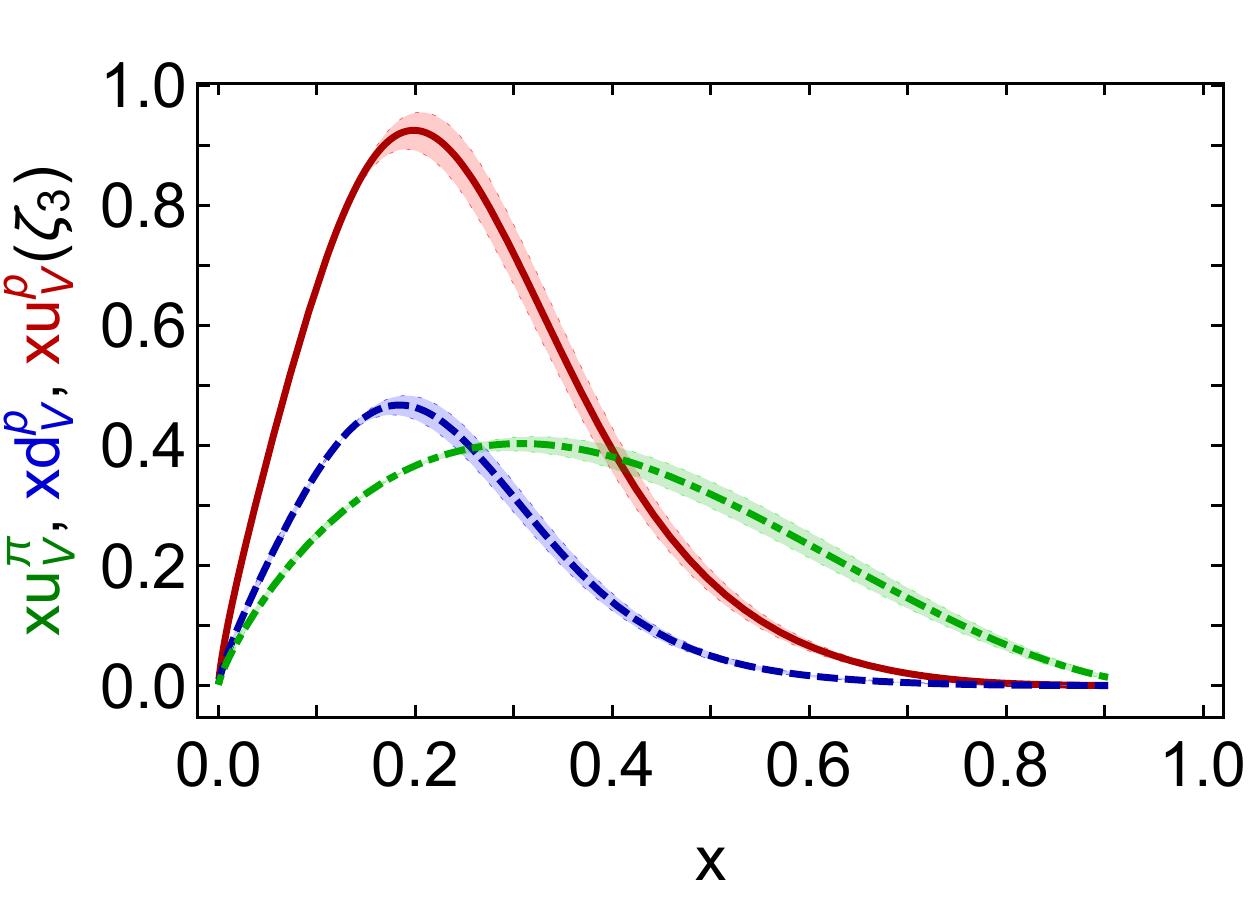}

\leftline{\hspace*{0.5em}{\large{\textsf{C}}}}
\vspace*{-3ex}
\includegraphics[width=0.41\textwidth]{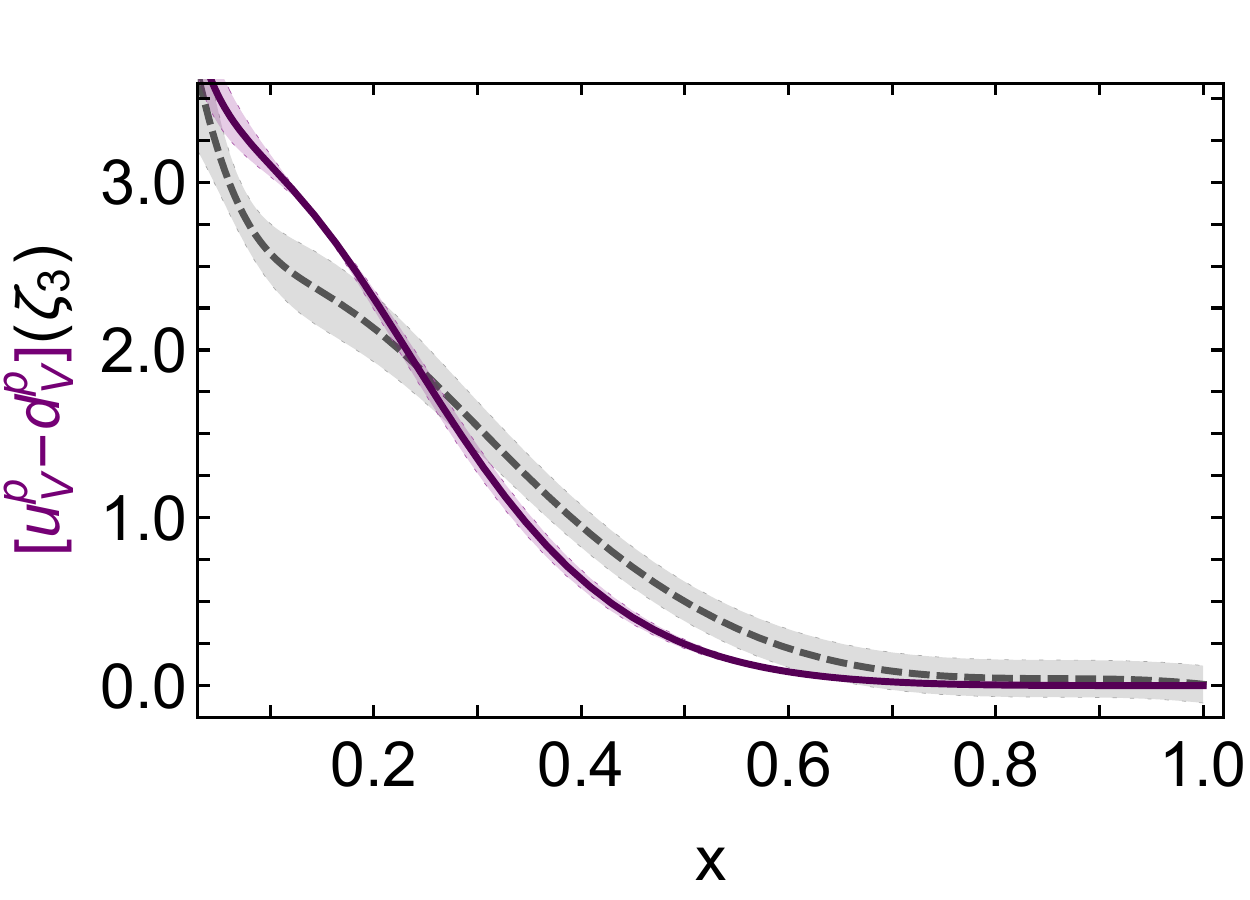}
\caption{\label{ImageValence}
\emph{Upper panel}\,--\,{\sf A}.
Hadron scale valence parton DFs for the proton and pion:
$x {\mathpzc u}^p(x;\zeta_{\cal H})$ -- solid red curve;
$x {\mathpzc d}^p(x;\zeta_{\cal H})$ -- dot-dashed blue curve;
and $x {\mathpzc u}^\pi(x;\zeta_{\cal H})$ -- dashed green curve.
\emph{Middle panel}\,--\,{\sf B}.
Valence DFs in panel {\sf A} evolved to $\zeta_3=m_{J/\psi}=3.097\,$GeV.
\emph{Lower panel}\,--\,{\sf C}.
Isovector distribution $[{\mathpzc u}^p(x;\zeta_{3}) - {\mathpzc d}^p(x;\zeta_{3})]$ (solid purple curve) compared with a lQCD result from Ref.\,\cite{Lin:2020fsj} (dashed grey curve).
The band surrounding each CSM curve expresses the response to a $\pm 5$\% variation in $\zeta_{\cal H}$.
}
\end{figure}

We subsequently work with the integro-differential evolution equations from which Eqs.\,\eqref{EqEvolution} are derived because they are satisfied by the DFs themselves and directly yield their $x$-dependence.  Their forms are obtained by using the PI charge described in Ref.\,\cite[Sec.\,3]{Cui:2020tdf} to integrate the one-loop DGLAP equations; and
\begin{equation}
\label{MassDependent}
 {\cal P}_{qg}^\zeta = \tfrac{1}{2} \left(1+\tanh[ (\zeta^2 - \delta_q^2)/\zeta_{\cal H}^2] \right)\,.
\end{equation}
Notably, as explained elsewhere \cite{Cui:2021sfu, Cui:2020tdf, Cui:2021mom, Cui:2022bxn}, the value of the hadron scale is a prediction: $\zeta_{\cal H} = 0.331(2)\,$.  Nevertheless, we report results with $\zeta_{\cal H} \to \zeta_{\cal H} (1\pm 0.05)$ in order to provide a conservative indication of uncertainty.

\medskip

\noindent\textbf{3.$\;$DFs at \mbox{\boldmath $\zeta = m_{J/\psi}$}}.
Modern CSM analyses of hadron scale valence DFs for the pion and proton are detailed elsewhere \cite{Cui:2020tdf, Chang:2022jri} and the results therein are drawn in Fig.\,\ref{ImageValence}A.
In considering these DFs, the following remarks are worth recording.
(\emph{a}) Each DF is consistent with the appropriate large-$x$ scaling law in Eq.\,\eqref{LargeX}.  Hence, from the beginning, whilst the $\zeta=\zeta_{\cal H}$ momentum sum rules are saturated by valence degrees-of-freedom for each hadron, Eqs.\,\eqref{MomSum} --
\begin{equation}
\langle x \rangle_{{\mathpzc u}_p}^{\zeta_{\cal H}}=0.687\,,\;
\langle x \rangle_{{\mathpzc d}_p}^{\zeta_{\cal H}} = 0.313\,,\;
\langle x \rangle_{{\mathpzc u}_\pi}^{\zeta_{\cal H}} =0.5\,,
\end{equation}
the pion and proton valence DFs nevertheless have markedly different pointwise behaviour.
(\emph{b}) Owing to DCSB \cite{Lane:1974he, Politzer:1976tv, Pagels:1978ba, Higashijima:1983gx, Roberts:2000aa, Binosi:2016wcx}, a corollary of EHM, QCD interactions simultaneously produce a dressed ${\mathpzc l}$-quark mass function, $M_{\mathpzc l}(k^2)$, that is large at infrared momenta, $M_D:= M_{\mathpzc l}(k^2\simeq 0)  \approx \zeta_{\cal H}$ and a nearly massless pion, $m_\pi^2/M^2_D \lesssim 0.2$.  (See the discussion in Ref.\,\cite[Sec.\,2]{Roberts:2021nhw}.)  Consequently, ${\mathpzc u}^\pi(x;\zeta_H)$ is Nature's most dilated hadron-scale valence DF.  This is exemplified in Fig.\,\ref{ImageValence}A and Refs.\,\cite{Cui:2021sfu, Cui:2020tdf}, and implicit in many other symmetry-preserving analyses, \emph{e.g}., Refs.\,\cite{Gao:2014bca, Binosi:2018rht, Ding:2018xwy, Lu:2021sgg}.


Employing the evolution scheme described in Sect.\,2, one obtains the $\zeta = m_{J/\psi}=:\zeta_3$ DFs in Fig.\,\ref{ImageValence}B.  Plainly, the individual valence degrees-of-freedom in the pion possess significantly more support on the valence domain than those in the proton.  This feature is an observable expression of EHM.

\begin{table}[t]
\caption{
\label{fitcoefficients}
Used in Eq.\,\eqref{EqInterpolation}, the listed powers and coefficients provide useful interpolations of all $\zeta=\zeta_3$ DFs calculated herein.
For the endpoint powers, $\alpha$, $\beta$, uncertainties associated with $\zeta_{\cal H}\to \zeta_{\cal H}(1\pm 0.05)$ are also shown.
}
\begin{tabular*}
{\hsize}
{
l@{\extracolsep{0ptplus1fil}}
c@{\extracolsep{0ptplus1fil}}
c@{\extracolsep{0ptplus1fil}}
c@{\extracolsep{0ptplus1fil}}
c@{\extracolsep{0ptplus1fil}}
c@{\extracolsep{0ptplus1fil}}
c@{\extracolsep{0ptplus1fil}}
c@{\extracolsep{0ptplus1fil}}
c@{\extracolsep{0ptplus1fil}}}\hline
%
$\pi$ & $\alpha\ $ & $\beta\ $ & $n_0\ $ & $n_1\ $ & $n_2\ $ & $d_1\ $ & $d_2\ $\\\hline
${\mathpzc u}\ $ & $\phantom{-}0.78_{(\pm 1)}\ $ & $2.47_{(\mp 7)}\ $ & $1.56\phantom{00}\ $ & $35.7\phantom{000}\ $ & $26.6\phantom{00}\ $ & $ 18.7\phantom{00}\ $& $-7.34\ $ \\
${\mathpzc g}\ $ & $-0.58_{(\pm 2)}\ $ & $3.88_{(\mp 7)}\ $ & $0.43\phantom{00}\ $ & $2.70\phantom{0}\ $ & $0.51\phantom{0}\ $ & $9.46\ $ & $-6.15\ $\\
${\mathpzc S}_u\ $ & $-0.49_{(\pm 2)}\ $ & $4.90_{(\mp8)}\ $ & $0.058\phantom{0}\ $ & $0.12\phantom{0}\ $ & $0.10\phantom{0}\ $& $5.00\ $& $-2.97\ $\\
${\mathpzc S}_s\ $ & $-0.51_{(\pm 2)}\ $  & $4.90_{(\mp8)}\ $ & $0.045\phantom{0}\ $ & $0.092\ $ & $0.081\ $ & $5.10\ $ & $-2.94\ $ \\
${\mathpzc S}_c\ $ &$-0.56_{(\pm 2)}\ $ & $4.96_{(\mp 8)}\ $  & $ 0.023\phantom{0}\ $ & $0.072\ $ & $0.024\ $ & $7.21\ $ & $-4.68\ $ \\\hline
\end{tabular*}

\medskip

\begin{tabular*}
{\hsize}
{
l@{\extracolsep{0ptplus1fil}}
c@{\extracolsep{0ptplus1fil}}
c@{\extracolsep{0ptplus1fil}}
c@{\extracolsep{0ptplus1fil}}
c@{\extracolsep{0ptplus1fil}}
c@{\extracolsep{0ptplus1fil}}
c@{\extracolsep{0ptplus1fil}}
c@{\extracolsep{0ptplus1fil}}
c@{\extracolsep{0ptplus1fil}}}\hline
$p$ & $\alpha\ $ & $\beta\ $ & $n_0\ $ & $n_1\ $ & $n_2\ $ & $d_1\ $ & $d_2\ $\\\hline
 ${\mathpzc u}\ $ & $\phantom{-}0.78_{(\pm 1)}\ $ & $4.11_{(\mp 6)}\ $ & $3.75\phantom{0}\ $ & $\phantom{-}0.79\phantom{0}\ $ & $20.7\phantom{500}\ $ & $-4.56\phantom{0}\ $ & $12.3\phantom{0}\ $\\
 ${\mathpzc d}\ $ & $\phantom{-}0.78_{(\pm 1)}\ $ & $4.11_{(\mp 6)}\ $  & $2.02\phantom{0}\ $ & $-1.47\phantom{0}\ $ & $4.88\phantom{0}\ $ & $-5.29\phantom{0}\ $ & $13.1\phantom{0}\ $\\
 ${\mathpzc g}\ $ & $-0.59_{(\pm 2)}\ $ &$5.45_{(\mp 6)}\ $  & $0.46\phantom{0}\ $ & $-0.93\phantom{0}\ $ & $0.76\phantom{0}\ $ & $-1.01\phantom{0}\ $ & $\phantom{1}1.63\ $\\
${\mathpzc S}_u\ $ & $-0.51_{(\pm 2)}\ $ & $6.41_{(\mp 6)}\ $ & $0.063\ $ & $-0.098\ $ & $0.055\ $& $\phantom{-}3.78\phantom{0}\ $& $-2.82\ $\\
${\mathpzc S}_d\ $ & $-0.51_{(\pm 2)}\ $ & $6.41_{(\mp 6)}\ $ & $0.069\ $ & $-0.12\phantom{0}\ $ & $0.12\phantom{0}\ $& $-0.022\ $& $\phantom{-}4.32\ $\\
${\mathpzc S}_s\ $ & $-0.52_{(\pm 2)}\ $ & $6.41_{(\mp 6)}\ $ & $0.051\ $ & $-0.084\ $ & $0.062\ $& $\phantom{-}1.64\phantom{0}\ $& $\phantom{-}0.30\ $\\
${\mathpzc S}_c\ $ & $-0.57_{(\pm 2)}\ $ & $6.41_{(\mp 5)}\ $ & $0.025\ $ & $-0.040\ $ & $0.025\ $& $\phantom{-}2.39\phantom{0}\ $& $-1.04\ $\\\hline
\end{tabular*}
\end{table}

The curves in Fig.\,\ref{ImageValence}B can usefully be interpolated using the following functional form:
\begin{equation}
\label{EqInterpolation}
x {\mathpzc p}(x) = x^\alpha (1-x)^\beta\,
\frac{n_0 + n_1 x + n_2 x^2}{1+d_1 x + d_2 x^2}\,,
\end{equation}
with the powers and coefficients listed in Table~\ref{fitcoefficients}.
(The powers are \emph{measurable} effective exponents, extracted from separate linear fits to $\ln[ x{\mathpzc p}(x)]$ on the domains $0<x<0.005$, $0.85<x<1$.)
Evidently, the valence distributions in the proton and pion each have the same power-law behaviour on $x\simeq 0$; and on $x\simeq 1$, $\beta_{\rm proton} \approx \beta_{\rm pion} + 1.6$, \emph{viz}.\ evolution to $\zeta > \zeta_{\cal H}$ preserves the differences in large-$x$ scaling behaviour described in Eq.\,\eqref{LargeX}.
Typical phenomenological fits to relevant scattering data yield DFs that fail to meet many of these QCD-based expectations, \emph{e.g}., Refs.\,\cite{Accardi:2016qay, NNPDF:2017mvq, Hou:2019efy}, something which increases the value of our predictions.  Additional discussion is provided elsewhere \cite{Courtoy:2020fex, Cui:2021mom, Cui:2022bxn}.

\begin{figure}[!t]
\vspace*{0.5ex}

\leftline{\hspace*{0.5em}{\large{\textsf{A}}}}
\vspace*{-3ex}
\includegraphics[width=0.41\textwidth]{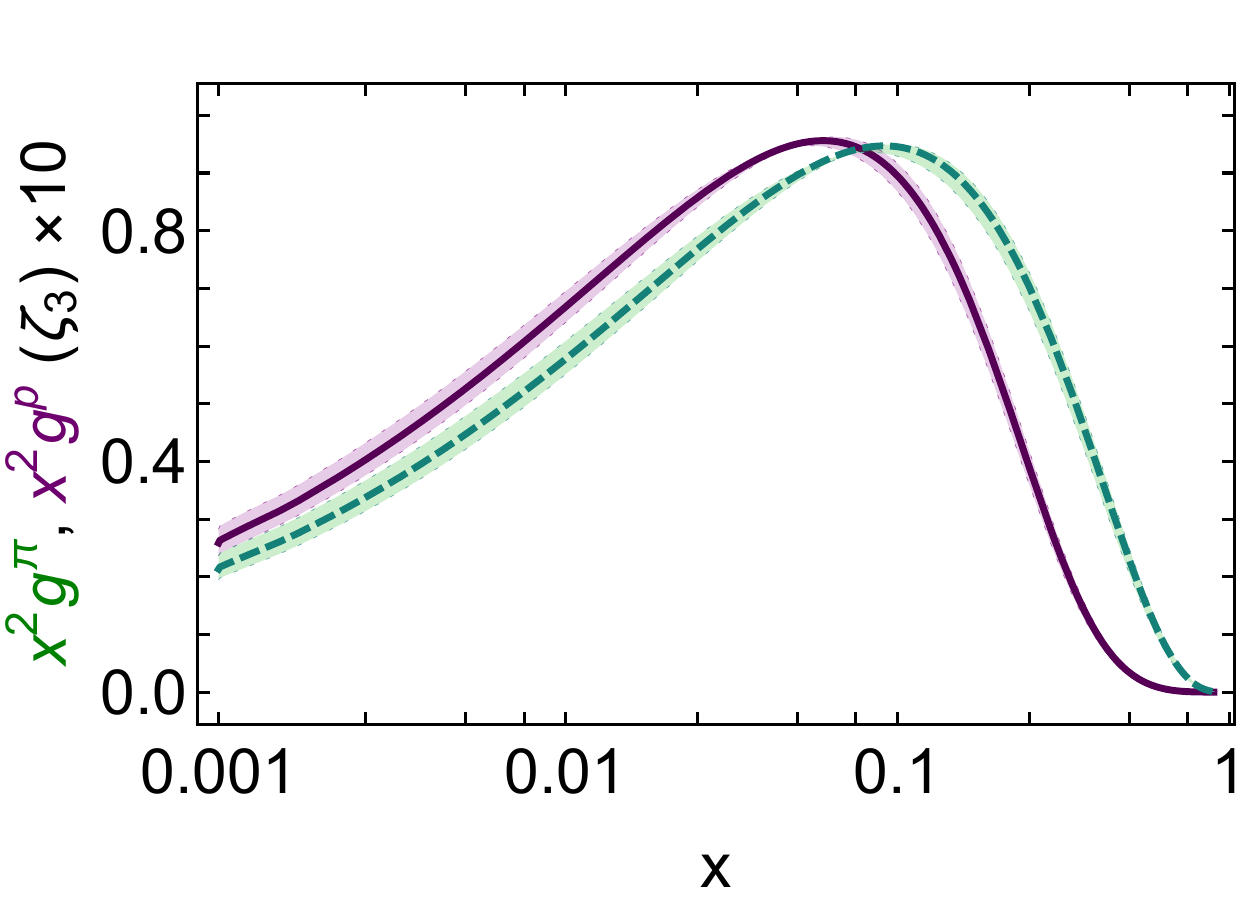}
\vspace*{-1ex}

\leftline{\hspace*{0.5em}{\large{\textsf{B}}}}
\vspace*{-3ex}
\includegraphics[width=0.41\textwidth]{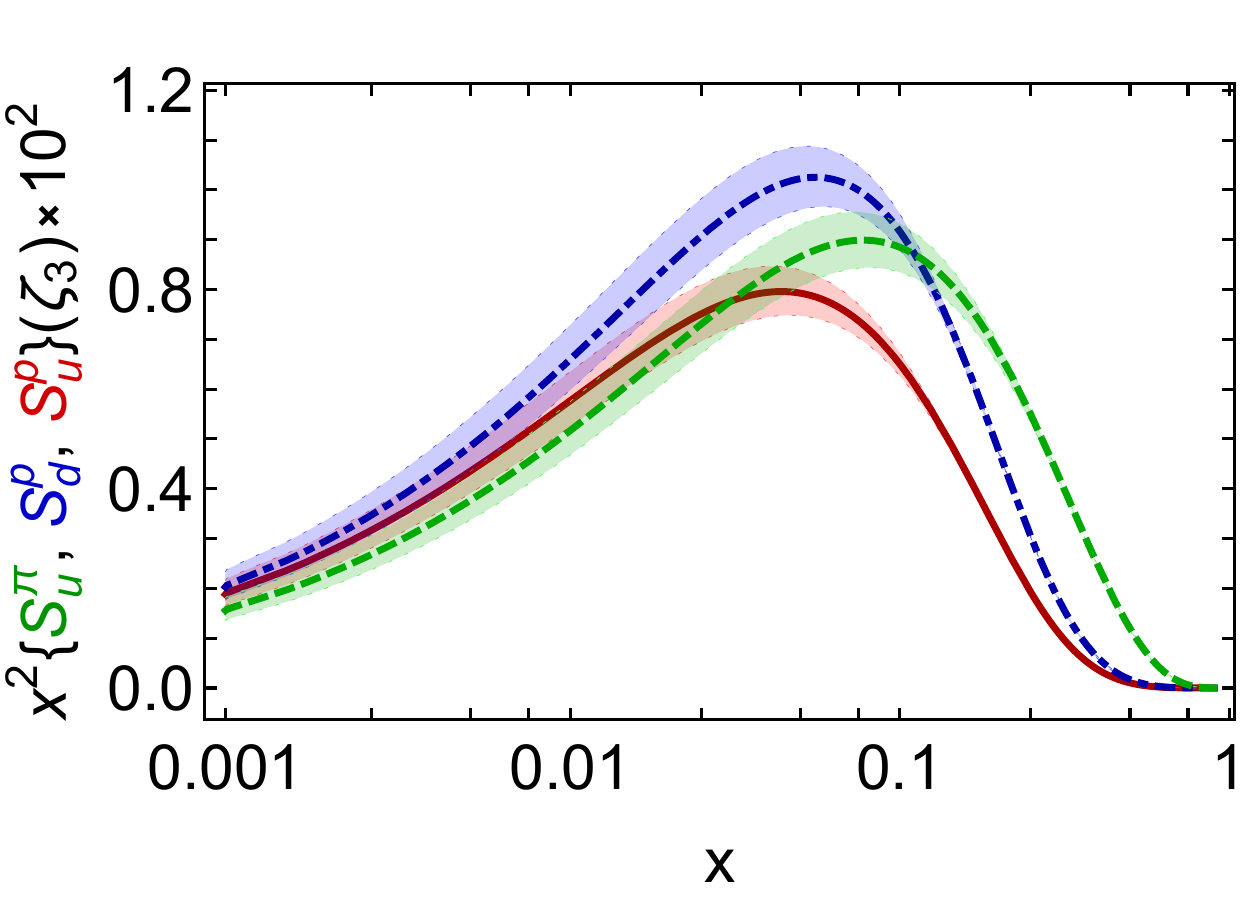}
\vspace*{-1ex}

\leftline{\hspace*{0.5em}{\large{\textsf{C}}}}
\vspace*{-3ex}
\includegraphics[width=0.41\textwidth]{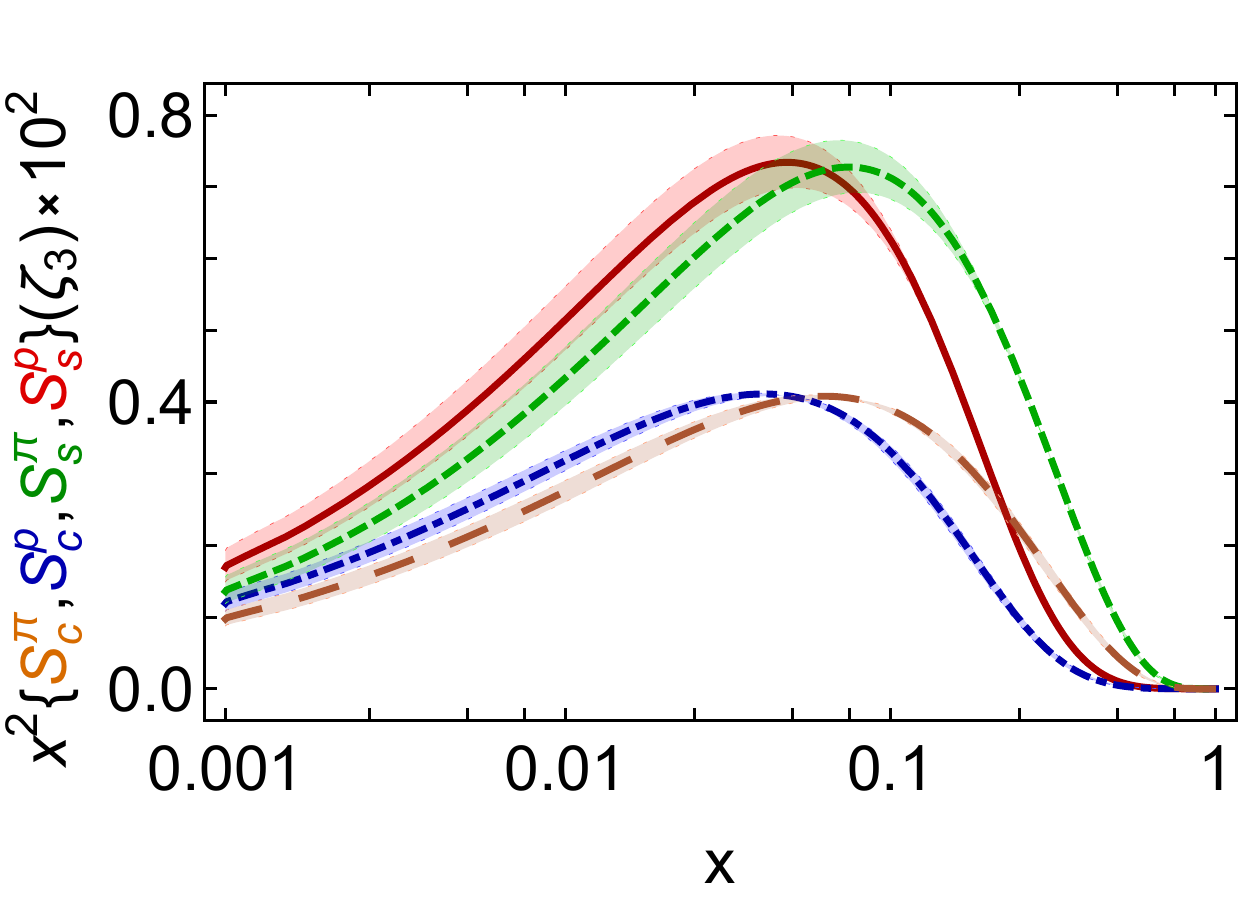}
\caption{\label{ImageGlue}
\emph{Upper panel}\,--\,{\sf A}.
Glue DFs -- $x^2 {\mathpzc g}$, in the proton (solid purple curve) and pion (dashed green curve) at $\zeta=\zeta_3$.  
\emph{Middle panel}\,--\,{\sf B}.
Light quark sea DFs in the proton and pion:
$x^2 {\mathpzc S}_u^p(x;\zeta_{3})$ -- solid red curve;
$x^2 {\mathpzc S}_d^p(x;\zeta_{3})$ -- dashed blue curve;
and $x^2 {\mathpzc S}_u^\pi(x;\zeta_{3})$ -- dot-dashed green curve.
\emph{Lower panel}\,--\,{\sf C}.
$c$- and $s$-quark sea DFs in the proton and pion:
$x^2 {\mathpzc S}_s^p(x;\zeta_{3})$ -- solid red curve;
$x^2 {\mathpzc S}_\pi^p(x;\zeta_{3})$ -- dashed green curve;
$x^2 {\mathpzc S}_c^p(x;\zeta_{3})$ -- dot-dashed blue curve;
and
$x^2 {\mathpzc S}_c^\pi(x;\zeta_{3})$ -- long-dashed orange curve.
(The band surrounding each curve expresses the response to a $\pm 5$\% variation in $\zeta_{\cal H}$.)
}
\end{figure}

\begin{figure}[!t]
\vspace*{0.5ex}

\leftline{\hspace*{0.5em}{\large{\textsf{A}}}}
\vspace*{-3ex}
\includegraphics[width=0.41\textwidth]{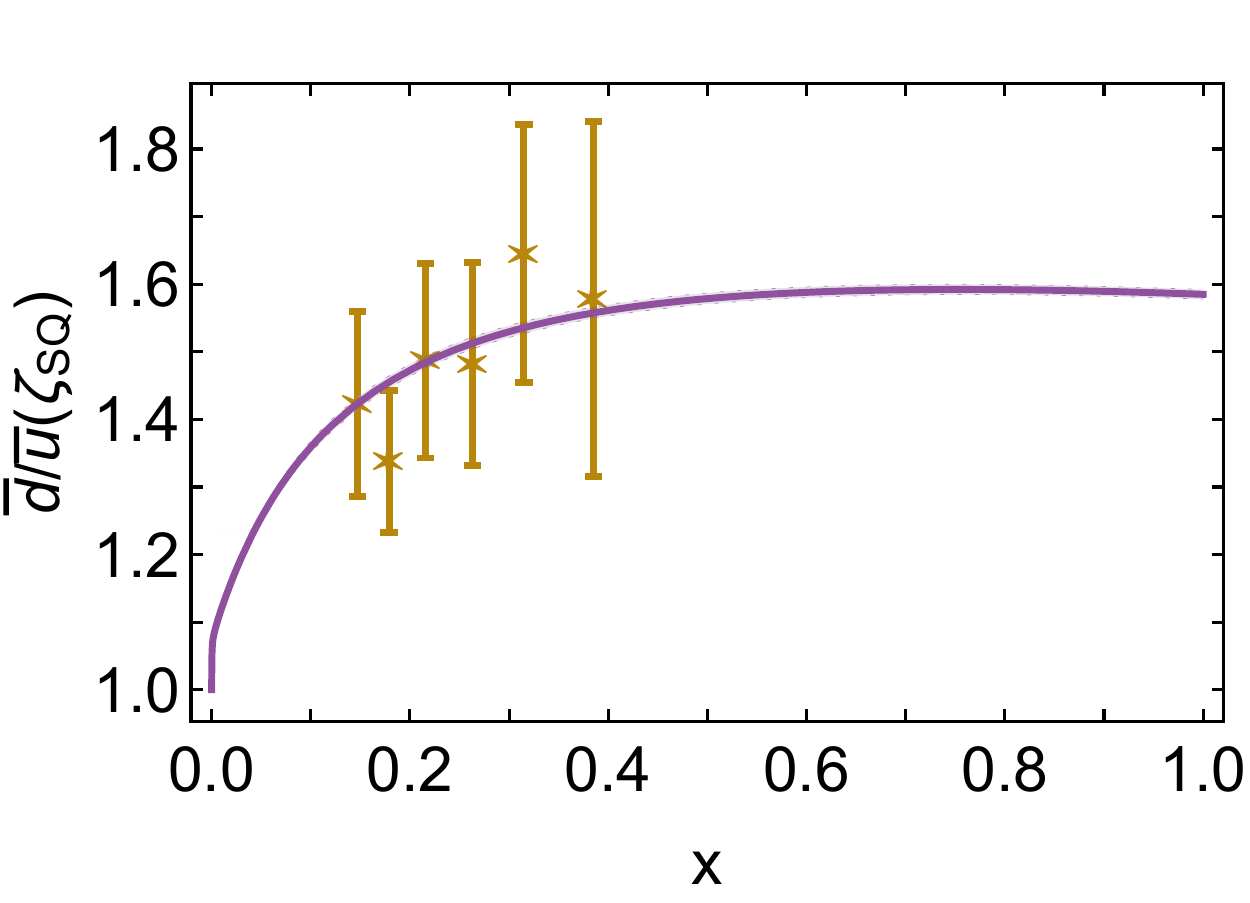}
\vspace*{-1ex}

\leftline{\hspace*{0.5em}{\large{\textsf{B}}}}
\vspace*{-3ex}
\includegraphics[width=0.41\textwidth]{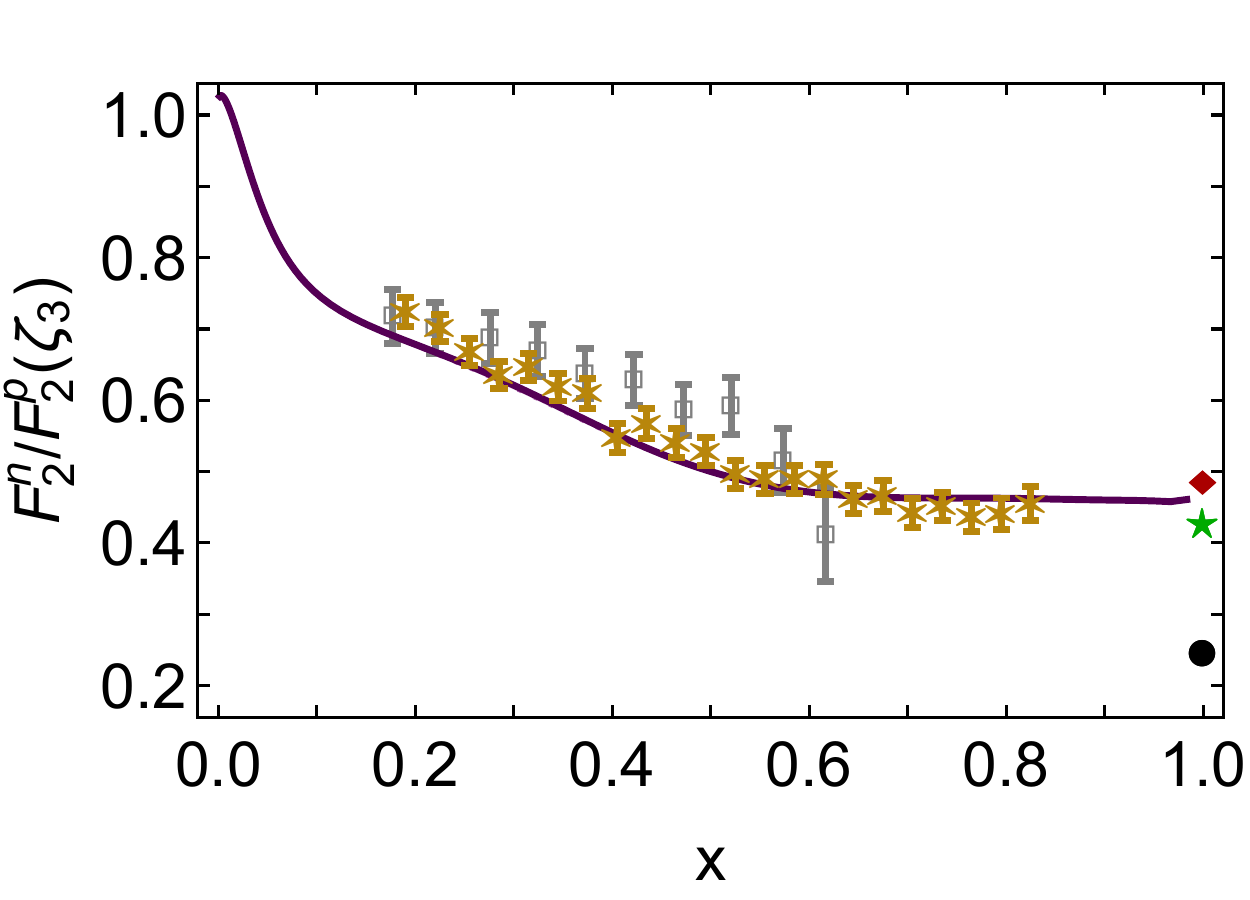}
\caption{\label{ImageSeaQuest}
\emph{Upper panel}\,--\,{\sf A}.
Ratio of light antiquark DFs.
Data from Ref.\,\cite[E906]{SeaQuest:2021zxb}.
Solid purple curve: result obtained from the valence-quark DFs in Fig.\,\ref{ImageValence}B after evolution to $\zeta^2=\zeta_{\rm SQ}^2 = 30\,$GeV$^2$.
\emph{Lower panel}\,--\,{\sf B}.
Neutron-to-proton structure function ratio.
Data: open grey squares \cite[BoNuS]{CLAS:2014jvt}; and gold asterisks \cite[MARATHON]{Abrams:2021xum}.
Solid purple curve: result obtained from valence-quark DFs in Fig.\,\ref{ImageValence}B after evolution to $\zeta=\zeta_{3}$.
Other predictions:
green star -- helicity conservation in the QCD parton model \cite{Farrar:1975yb, Brodsky:1979gy, Brodsky:1994kg};
red diamond -- continuum Schwinger function methods \cite{Roberts:2013mja};
and retaining only scalar diquarks in the proton wave function, which produces a large-$x$ value for this ratio that lies in the neighbourhood of the filled circle \cite{Close:1988br, Xu:2015kta}.
(Both panels: narrow band bracketing each curve expresses response to $\pm 5$\% variation in $\zeta_{\cal H}$).
}
\end{figure}

Owing to difficulties in handling so-called disconnected contributions, the calculation of individual valence DFs using lattice-regularised QCD (lQCD) is problematic \cite{Alexandrou:2013cda}; so, lQCD results are typically only available for isovector distributions, from which disconnected contributions vanish in the continuum limit.  Therefore, Fig.\,\ref{ImageValence}C displays the isovector distribution $[{\mathpzc u}^p(x;\zeta_{3}) - {\mathpzc d}^p(x;\zeta_{3})]$, calculated from the curves in Fig.\,\ref{ImageValence}B, along with a lQCD result from Ref.\,\cite{Lin:2020fsj}, extracted using large-momentum effective theory and extrapolated to a continuum limit and physical pion mass.  The level of agreement is encouraging, especially because refinements of both calculations may be anticipated.

When evolving singlet and glue DFs, we include a Pauli blocking factor in the gluon splitting function, as discussed elsewhere \cite[Sec.\,6]{Chang:2022jri}:
\begin{equation}
\label{gluonsplit}
P_{f \leftarrow g}(x;\zeta) \to P_{f \leftarrow g}(x) +
 \sqrt{3}  (1 - 2 x) \frac{ {\mathpzc g}_{f} }{1+(\zeta/\zeta_H-1)^2}\,,
\end{equation}
where $P_{f \leftarrow g}(x) $ is the standard one-loop gluon splitting function, ${\mathpzc g}_{s,\bar s}=0={\mathpzc g}_{c,\bar c}$, and ${\mathpzc g}_{d,\bar d}= 0.34 =  -{\mathpzc g}_{u,\bar u}=:{\mathpzc g}$ is a strength parameter.
This term preserves baryon number.  It shifts momentum into $d+\bar d$ from $u+\bar u$, otherwise leaving the total sea momentum fraction unchanged, and vanishes with increasing $\zeta$, reflecting the waning influence of valence-quarks as the proton's glue and sea content increases.

Our predictions for the $\zeta=\zeta_3$ glue DFs in the proton and pion are drawn in Fig.\,\ref{ImageGlue}A.  Regarding the glue-in-$\pi$ DF, our result is straightforwardly connected via evolution to the form in Ref.\,\cite{Chang:2021utv}; thus, as discussed therein, it agrees with a recent lQCD calculation of this DF \cite{Fan:2021bcr}.
It is clear from Fig.\,\ref{ImageGlue}A that the glue-in-$\pi$ DF possess significantly more support on the valence domain than the kindred glue-in-$p$ DF.  This outcome is also a measurable expression of EHM.

\begin{table*}[!t]
\caption{
\label{TabMoments}
Low-order Mellin moments, $\langle x^m\rangle_{{\mathpzc p}_H}^{\zeta_3}$, of the DFs drawn in Figs.\,\ref{ImageValence}B\,--\,\ref{ImageGlue}, measured in \%.
As an illustration of the numerical accuracy of our evolution procedure, we note that $\langle x\rangle_{{\mathpzc c}_\pi}^{\zeta_3}$ and $\langle x\rangle_{{\mathpzc c}_p}^{\zeta_3}$ differ by only 0.3\%.
Uncertainties associated with $\zeta_{\cal H}\to \zeta_{\cal H}(1\pm 0.05)$ are shown.
To simplify comparisons with phenomenological fits to relevant data, results for $\langle x^m\rangle_{{\mathpzc p}_H}^{\zeta_2}$, $\zeta_2=2\,$GeV, are also listed.
The $m=1,2,3$ moments of the proton isovector distribution, $[u-d]$, are: $\zeta_2$ -- $17.9(8)$\%, $5.1(3)$\%, $1.8(2)$\%; and $\zeta_3$ -- $16.6(7)$\%, $4.5(3)$\%, $1.6(1)$\%
}
\begin{tabular*}
{\hsize}
{
l@{\extracolsep{0ptplus1fil}}
c@{\extracolsep{0ptplus1fil}}
c@{\extracolsep{0ptplus1fil}}
c@{\extracolsep{0ptplus1fil}}
c@{\extracolsep{0ptplus1fil}}
c@{\extracolsep{0ptplus1fil}}
c@{\extracolsep{0ptplus1fil}}
c@{\extracolsep{0ptplus1fil}}}\hline
%
pion$\rule{0em}{2.5ex}\ $ & ${\mathpzc u}^\pi \ $ & $\bar{\mathpzc d}^\pi \ $ & ${\mathpzc g}^\pi\ $ &
${\mathpzc S}_\pi^u\ $ & ${\mathpzc S}_\pi^{\bar d}\ $ & ${\mathpzc S}_\pi^s\ $ & ${\mathpzc S}_\pi^c\ $ \\ \hline
$\langle x \rangle^{\zeta_2}\ $ & $24.0(1.1) $ & $24.0(1.1)\ $ & $41.0(1.2)\ $& $3.3(3)\ $ & $3.3(3)\ $& $2.65(22)\ $ & $1.33(5)\ $ \\
$\langle x^2\rangle^{\zeta_2}\ $ & $9.5(7)\ $ & $9.5(7)\ $ & $3.7(1)\ $& $\phantom{1}0.27(1)\ $ & $\phantom{1}0.27(1)\ $ & $\phantom{1}0.21(1)\phantom{11}\ $ &$\phantom{1}0.092(2)\ $ \\
$\langle x^3\rangle^{\zeta_2}\ $ & $4.7(4)\ $ & $4.7(4)\ $ & $\phantom{1}0.92(6)\ $& $\phantom{11}0.057(1)\ $ & $\phantom{11}0.057(1)\ $ & $\phantom{11}0.044(0)\phantom{11}\ $ & $\phantom{1}0.018(1)\ $\\
\hline
$\langle x \rangle^{\zeta_3}\ $ & $22.1(1.0) $ & $22.1(1.0)\ $ & $42.9(1.0)\ $& $3.7(3)\ $ & $3.7(3)\ $& $3.0(2)\phantom{11}\ $ & $1.83(6)\ $ \\
$\langle x^2\rangle^{\zeta_3}\ $ & $8.4(6)\ $ & $8.4(6)\ $ & $3.5(1)\ $& $\phantom{1}0.27(1)\ $ & $\phantom{1}0.27(1)\ $ & $\phantom{1}0.22(1)\phantom{11}\ $ &$\phantom{1}0.120(3)\ $ \\
$\langle x^3\rangle^{\zeta_3}\ $ & $4.0(3)\ $ & $4.0(3)\ $ & $\phantom{1}0.82(5)\ $& $\phantom{11}0.056(0)\ $ & $\phantom{11}0.056(0)\ $ & $\phantom{11}0.044(0)\phantom{11}\ $ & $\phantom{1}0.022(1)\ $\\
\hline
\end{tabular*}

\medskip

\begin{tabular*}
{\hsize}
{
c@{\extracolsep{0ptplus1fil}}
c@{\extracolsep{0ptplus1fil}}
c@{\extracolsep{0ptplus1fil}}
c@{\extracolsep{0ptplus1fil}}
c@{\extracolsep{0ptplus1fil}}
c@{\extracolsep{0ptplus1fil}}
c@{\extracolsep{0ptplus1fil}}
c@{\extracolsep{0ptplus1fil}}}\hline
%
proton & ${\mathpzc u}^p \ $ & ${\mathpzc d}^p \ $ & ${\mathpzc g}^p\ $ &
${\mathpzc S}_p^u\ $ & ${\mathpzc S}_p^{d}\ $ & ${\mathpzc S}_p^s\ $ & ${\mathpzc S}_p^c\ $ \\ \hline
%
$\langle x \rangle^{\zeta_2}\ $ & $32.9(1.4) $ & $15.0(0.7)\ $ & $40.9(1.1)\ $& $2.9(2)\ $ & $3.7(3)\ $ & $2.64(22)\ $& $1.32(5)\ $ \\
$\langle x^2\rangle^{\zeta_2}\ $ & $8.7(6)\ $ & $3.6(2)\ $ & $2.4(1)\ $& $\phantom{1}0.14(1)\ $& $\phantom{1}0.21(1)\ $ & $\phantom{1}0.13(0)\phantom{11}\ $ & $\phantom{1}0.059(2)\ $\\
$\langle x^3\rangle^{\zeta_2}\ $ & $2.9(3)\ $ & $1.1(1)\ $ & $\phantom{1}0.39(2)\ $& $\phantom{11}0.019(0)\ $ & $\phantom{11}0.030(1)\ $ & $\phantom{11}0.019(0)\phantom{11}\ $ & $\phantom{1}0.008(0)\ $ \\
\hline
$\langle x \rangle^{\zeta_3}\ $ & $30.4(1.3) $ & $13.8(0.6)\ $ & $42.8(1.0)\ $& $3.3(3)\ $ & $4.1(3)\ $ & $3.0(2)\phantom{11}\ $& $1.82(6)\ $ \\
$\langle x^2\rangle^{\zeta_3}\ $ & $7.7(5)\ $ & $3.2(2)\ $ & $2.2(1)\ $& $\phantom{1}0.15(1)\ $& $\phantom{1}0.21(1)\ $ & $\phantom{1}0.14(0)\phantom{11}\ $ & $\phantom{1}0.075(2)\ $\\
$\langle x^3\rangle^{\zeta_3}\ $ & $2.5(2)\ $ & $0.9(1)\ $ & $\phantom{1}0.35(2)\ $& $\phantom{11}0.019(0)\ $ & $\phantom{11}0.028(0)\ $ & $\phantom{11}0.019(0)\phantom{11}\ $ & $\phantom{1}0.010(1)\ $ \\
\hline
\end{tabular*}
\end{table*}

Useful interpolations of the curves in Fig.\,\ref{ImageGlue}A are obtained using Eq.\,\eqref{EqInterpolation} and the relevant powers and coefficients in Table~\ref{fitcoefficients}.
The powers are interesting.  On $x\simeq 0$, the proton and pion glue DFs exhibit practically the same power-law growth; and on $x\simeq 1$, confirming the QCD expectations reported in connection with Eq.\,\eqref{LargeX}, $\beta_{\rm glue} \approx \beta_{\rm valence} + 1.4$ for both proton and pion.
The endpoint exponents on glue-in-$p$ DFs are discussed in Ref.\,\cite{Sufian:2020wcv}, from a lQCD perspective within the context of Ioffe-time distributions.  Lattice-QCD computations are currently insensitive to low-$x$ physics.  On the other hand, a meaningful estimate of the large-$x$ exponent is reported \cite{Sufian:2020wcv}: $\beta(\zeta=2\,{\rm GeV}) = 4.9(1.2)$.  Our approach delivers the following $\zeta=\zeta_2$ values: $\alpha_{\rm proton}^{\rm glue} = -0.56(2) $, $\beta_{\rm proton}^{\rm glue} = 5.33(5)$ and $\alpha_{\rm pion}^{\rm glue} = 0.54(2)$, $\beta_{\rm pion}^{\rm glue} = 3.75(5)$.

Evolving the valence DFs in Fig.\,\ref{ImageValence}A to $\zeta=\zeta_3$, one obtains the light-quark sea DFs for the proton and pion depicted in Fig.\,\ref{ImageGlue}B.  In keeping with the EHM-induced pattern already established, the sea-in-$\pi$ DF possess significantly more support on the valence domain than the kindred sea-in-$p$ DFs.

Interpolations of the curves in Fig.\,\ref{ImageGlue}B are provided by Eq.\,\eqref{EqInterpolation} and the relevant powers and coefficients in Table~\ref{fitcoefficients}.  Once again, the low- and high-$x$ exponents match QCD expectations: on $x\simeq 0$, the proton and pion light-sea DFs exhibit approximately the same power-law growth; and on $x\simeq 1$, $\beta_{\rm sea} \approx \beta_{\rm valence} + 2.4$ for both proton and pion.

Owing to the Pauli blocking factor, Eq.\,\eqref{gluonsplit}, an in-proton separation between $\bar d$ and $\bar u$  is evident in Fig.\,\ref{ImageGlue}B.  This entails a violation of the Gottfried sum rule \cite{Gottfried:1967kk, Brock:1993sz}, which has been found in a series of experiments \cite{NewMuon:1991hlj, NewMuon:1993oys, NA51:1994xrz, NuSea:2001idv, SeaQuest:2021zxb}.
Using the DFs in Fig.\,\ref{ImageGlue}B, one obtains
\begin{equation}
\label{gottfried}
\int_{0.004}^{0.8} dx\,[\bar {\mathpzc d}(x;\zeta_3) - \bar {\mathpzc u}(x;\zeta_3)]
= 0.116(12)
\end{equation}
for the Gottfried sum rule discrepancy on the domain covered by the measurements in Refs.\,\cite{NewMuon:1991hlj, NewMuon:1993oys}.  This value may be compared with that inferred from recent fits to a variety of high-precision data ($\zeta = 2\,$GeV) \cite[CT18]{Hou:2019efy}: 0.110(80).
%
Evolved to $\zeta_{\rm SQ}^2=30\,$GeV$^2$, the result in Eq.\,\eqref{gottfried} becomes $0.110(11)$, a value that is $\lesssim 20$\% larger than that determined in Ref.\,\cite{Chang:2022jri}, which ignored quark mass effects in the evolution equations.  On the other hand, we implemented mass-dependent evolution via Eq.\,\eqref{MassDependent} and this increases the magnitudes of the proton's light-quark sea DFs.
Nevertheless, as revealed by Fig.\,\ref{ImageSeaQuest}A, our result for the ratio $\bar {\mathpzc d}(x;\zeta_{\rm SQ})/\bar {\mathpzc u}(x;\zeta_{\rm SQ})$ reproduces that in Ref.\,\cite[Fig.\,2B]{Chang:2022jri} and matches modern data Ref.\,\cite[E906]{SeaQuest:2021zxb}.

Using the scheme described in Sect.\,2, DFs for heavier sea quarks are also generated via evolution. The predictions are drawn in Fig.\,\ref{ImageGlue}C.  Evidently, the $\zeta=\zeta_3$ $s$ and $c$ quark sea DFs are commensurate in size with those of the light-quark sea DFs; and, for $s$-and $c$-quarks, too, the pion DFs possess significantly greater support on the valence domain than the kindred proton DFs.
Interpolations of the curves in Fig.\,\ref{ImageGlue}C are provided by Eq.\,\eqref{EqInterpolation} and the appropriate powers and coefficients in Table~\ref{fitcoefficients}.  The low- and high-$x$ exponents match QCD expectations: on $x\simeq 0$, the proton and pion light-sea DFs exhibit very similar power-law growth; and on $x\simeq 1$, one also finds $\beta_{\rm sea} \approx \beta_{\rm valence} + 2.4$ for both proton and pion.

Using our results for the valence and sea DFs, it is straightforward to calculate the neutron-proton structure function ratio:
\begin{align}
\label{F2nF2p}
\frac{F_2^n(x;\zeta)}{F_2^p(x;\zeta)} =
\frac{
{\mathpzc U}(x;\zeta) + 4 {\mathpzc D}(x;\zeta) + \Sigma(x;\zeta)}
{4{\mathpzc U}(x;\zeta) + {\mathpzc D}(x;\zeta) + \Sigma(x;\zeta)}\,,
\end{align}
where, in terms of quark and antiquark DFs,
${\mathpzc U}(x;\zeta) = {\mathpzc u}(x;\zeta)+\bar {\mathpzc u}(x;\zeta)$,
${\mathpzc D}(x;\zeta) = {\mathpzc d}(x;\zeta)+\bar {\mathpzc d}(x;\zeta)$,
$\Sigma(x;\zeta) = {\mathpzc s}(x;\zeta)+\bar {\mathpzc s}(x;\zeta)
  +{\mathpzc c}(x;\zeta)+\bar {\mathpzc c}(x;\zeta)$.
The $\zeta=\zeta_3$ prediction is drawn in Fig.\,\ref{ImageSeaQuest}B: in comparison with modern data \cite[MARATHON]{Abrams:2021xum}, it yields $\chi^2/$degree-of-freedom$ \;=1.3$.  Notably, both data and calculation indicate the presence of a significant axial-vector diquark component in the proton wave function \cite{Barabanov:2020jvn, Cui:2021gzg}.

As remarked above, data on pion DFs is scarce and some controversy attends interpretations of such data \cite{Cui:2021mom}.  Notwithstanding these things, the pion DFs calculated herein are viable, as demonstrated elsewhere \cite{Cui:2021sfu, Cui:2020tdf, Chang:2021utv, Roberts:2021nhw}.

Low-order Mellin moments of all proton and pion DFs are listed in Table~\ref{TabMoments}.
As signalled above, our approach entails that comparable momentum fractions in the proton and pion are identical and the total sea-quark momentum fraction is shared amongst the quarks in roughly inverse proportion to their dressed-mass, $M_q$.
Importantly, the calculated values of the listed DF moments are in fair agreement with those computed from phenomenological fits obtained using a variety of methods; see, \emph{e.g}., Ref.\,\cite[Table~VI]{Hou:2019efy}: referred to the CT18 column, our results match at the level of $1.7(1.5)\,\sigma$.
This quantitative similarity also extends to the $c$ quark: we find
$\langle x\rangle_{{\mathpzc c}_p}^{\zeta_2}=1.32(5)$\%, $\langle x\rangle_{{\mathpzc c}_p}^{\zeta_3}=1.82(6)$\%,
which may respectively be compared with the values $1.7(4)$, $2.5(4)$\% in Ref.\,\cite[Fig.\,60]{NNPDF:2017mvq}.
Such an array of correspondences is noteworthy because our results are predictions, derived from the pion and proton wave functions in Refs.\,\cite{Cui:2021sfu, Cui:2020tdf, Chang:2022jri}, using only one free parameter, \emph{viz}.\ ${\mathpzc g}$ in Eq.\,\eqref{gluonsplit} to introduce an asymmetry of antimatter in the proton.

Potentially drawing a line to the notion of intrinsic charm \cite{Brodsky:1980pb}, it is worth highlighting that our approach yields $\langle x\rangle_{{\mathpzc c}}^{\zeta=M_c}=0.64(3)$\% in both the pion and proton.  Regarding the pion, nothing is known about this momentum fraction; and in the proton, phenomenological estimates are inconclusive, ranging from $0$-$2$\% \cite[Fig.\,59]{NNPDF:2017mvq}.
Notwithstanding the size of these calculated fractions, we stress that ${\mathpzc S}_{\pi,p}^c(x)$ have sea-quark profiles.

We have shown that contemporary CSM results for proton and pion $\zeta=\zeta_{\cal H}$ valence DFs, obtained from symmetry-preserving analyses and used as initial values for evolution according to proposition {\sf P1}, yield predictions for the pointwise behaviour of all proton and pion $\zeta > \zeta_{\cal H}$ DFs (valence, sea, glue) that are consistent with QCD expectations, including those described in connection with Eq.\,\eqref{LargeX}.
In contrast, extant phenomenological fits to relevant data are inconsistent with one or more of these constraints.  Consequently, such fits cannot serve as a reliable foundation for evaluating the validity of evolution schemes such as that described in Sect.\,2.  In large part, this explains conclusions drawn elsewhere \cite{Diehl:2019fsz}.  Future such studies should be built upon improved DF fits and use an effective charge that excludes a Landau pole and so furnishes an infrared completion of QCD.
%

\medskip

\noindent\textbf{4.$\;$Perspective}.
Beginning with hadron-scale proton and pion valence distribution functions (DFs) obtained using symmetry-preserving treatments of the continuum bound-state problem and assuming only that there is an effective charge which defines an evolution scheme for parton DFs that is all-orders exact, we delivered a unified body of predictions for all proton and pion DFs -- valence, glue, and four-flavour-separated sea.
Notably, within mesons and baryons that share a familial flavour structure, this evolution approach entails that the hadron light-front momentum fractions carried by identifiable, distinct parton classes are the same at any scale.  Notwithstanding that, providing a measurable expression of emergent hadron mass, the pointwise behaviour of the distributions is strongly hadron-dependent: at any resolving scale, $\zeta$, those in the pion are the hardest (most dilated).
The framework's viability was illustrated by comparisons with the $x$-dependence of modern data,
results from lattice-regularised QCD, and also Mellin moments computed using contemporary phenomenological DF fits.

Of particular significance is the result that all DFs calculated herein comply with QCD constraints on endpoint (low- and high-$x$) scaling behaviour.  In our view, only after imposing these constraints on future phenomenological fits to relevant scattering data will it be possible to draw reliable pictures of hadron structure.  This will be especially important for attempts to expose and understand the differences between Nambu-Goldstone bosons and seemingly less complex hadrons.

Although the Poincar\'e-covariant pion wave function used herein is sophisticated, having been validated through numerous applications, that of the proton is an \emph{Ansatz} informed by modern continuum Schwinger function analyses.   It is therefore worth repeating this study using a refined form.  One may also expect that, in the longer term, the analysis herein could be undertaken using direct solutions of a three-body Faddeev equation for the proton \cite{Eichmann:2009qa, Wang:2018kto}, raising the proton wave function to the same level as that of the pion.


\medskip
\noindent\textbf{Acknowledgments}.
We are grateful for constructive comments from D.~Binosi, C.~Chen, Z.-F.~Cui, M.~Ding, F.~Gao, R.~Sufian and S.\,M.~Schmidt.
Work supported by:
National Natural Science Foundation of China (grant no.\,12135007);
Spanish Ministry of Science and Innovation (MICINN) (grant no.\ PID2019-107844GB-C22);
and
Junta de Andaluc{\'{\i}}a (grant nos.\ P18-FR-5057, UHU-1264517).


\medskip
\noindent\textbf{Declaration of Competing Interest}.
The authors declare that they have no known competing financial interests or personal relationships that could have appeared to influence the work reported in this paper.



\end{document}